\documentclass{jfm}
\usepackage[process=auto,crop=pdfcrop, cleanup={.tex, .dvi, .ps, .pdf, .log, .bbl}]{pstool}
\usepackage{t1enc,latexsym} 
\usepackage[T1]{fontenc}
\usepackage{nicefrac}
\usepackage{amsmath}
\usepackage{booktabs}

\DeclareMathAlphabet{\mathsf}{OT1}{\sfdefault}{m}{n}
\SetMathAlphabet{\mathsf}{bold}{OT1}{\sfdefault}{b}{n}

\newcommand{\bVec}[1]{\mathbi{#1}}
\newcommand{\Ten}[1]{#1}
\newcommand{\bTen}[1]{\mathsfbi{#1}}
\newcommand{\Eq}[1]{Eq. (\ref{eq:#1})}

\newcommand{\fig}[1]{figure (\ref{fig:#1})}

\newcommand{\tab}[1]{table (\ref{tab:#1})}
\newcommand\Sc{\mbox{\textit{Sc}}}
\newcommand{\UGC}[0]{\mathrm R}
\newcommand{\Dbar}[0]{\text{\emph{\DJ}}}
\newcommand{\figwidth}[0]{\linewidth}

\newcommand{\InsPSfragFig}[3]{
	\begin{figure}
		\begin{center}
			\psfragfig[width=\figwidth]{#1}
			\caption{#2}
			\label{fig:#3}
		\end{center}
	\end{figure}
}

\shorttitle{On Diffusion-Induced Boundary-Layer Composition Profiles}
\shortauthor{S. G. Johnsen}

\title{On Diffusion-Induced Non-Constant Composition Profiles in the Boundary Layer of Inert Multicomponent Mixtures}

\author{Sverre Gullikstad Johnsen\aff{1}\aff{2}
  \corresp{\email{sverregu@gmail.com}}}

\affiliation{
  \aff{1}SINTEF Industry, Trondheim, Norway
  \aff{2}NTNU, dept. Materials Science and Engineering, Trondheim, Norway
}

\begin{document}
\maketitle

\begin{abstract}
In the boundary layer of multicomponent fluid mixtures, the species-specific mass flux in the wall-normal direction is determined by the combination of turbulent-diffusiophoretic diffusion due to composition gradients, and diffusion due to gradients in other scalar fields (e.g. thermophoresis, barophoresis, and forced diffusion).
For inert mixtures, a balance must exist between all the diffusive transport mechanisms so that the net diffusive mass flux normal to the wall is zero everywhere.
This paper discusses under which conditions non-constant composition profiles are necessary to obtain physico-chemical equilibrium, hence vanishing transport, in the wall-normal direction.

Mathematical modeling is employed to demonstrate how this may affect fluid property profiles, wall heat flux, and wall shear stress in an ideal, ternary gas mixture ($H_2 + N_2 + CO_2$) subject to a temperature gradient.

It is concluded that competing diffusive transport mechanisms, under certain circumstances, may result in non-constant composition profiles and significantly different wall and bulk compositions, for inert multicomponent mixtures.
Hence, wall-normal diffusion in the boundary layer must be accounted for to correctly describe or interpret multicomponent flow systems sensitive to wall friction and heat exchange.
\end{abstract}
%

\section{Introduction}
The classical experiment by \citet{Duncan1962} showed that the diffusive transport in an ideal ternary gas mixture of hydrogen ($H_2$), nitrogen ($N_2$), and carbon dioxide ($CO_2$) could not be described satisfactorily by Fickian diffusion theory.
E.g., the observed development in local nitrogen concentrations could only be explained mathematically by allowing uphill diffusion.
The Duncan-Toor experiments have been further investigated and discussed by e.g. \citet{Taylor93} and \citet{Krishna1997}.
It has been shown that Maxwell-Stefan diffusion theory predicts the non-Fickian behavior observed by Duncan and Toor, accurately.

Whereas the Duncan-Toor experiments were performed under isothermal conditions, \citet{Bogatyrev2015} studied thermophoresis in binary, ternary, and quaternary mixtures including the ternary $H_2-N_2-CO_2$ mixture.
They emphasized that thermophoresis in multicomponent mixtures depends on the mixture composition in a complex way.

\citet{Gibbs1928} was the first to suggest a formula for calculating compositional variations due to gravity, based on vanishing gradients in the physico-chemical potential.
Later, significant compositional gradients observed in subsurface oil/gas reservoirs have been credited gravity-chemical equilibrium, e.g. in the Brent field in the North Sea \citep{Schulte1980}.
Moreover, it is well known that centrifugal forces can induce species segregation \citep{Svedberg1927}.
This phenomenon has widespread industrial and scientific application in e.g. materials science, chemistry, and biotechnology.

Physico-chemical equilibrium demands zero net wall-normal transport for each species, everywhere in the boundary layer of inert mixtures.
It is evident, however, that competing diffusive processes which cancel each other out, can fulfill this requirement.
Hence, it is suggested that under certain conditions, non-constant composition profiles in the wall-normal direction may result from non-Fickian behavior.

Such spatial composition variations will affect fluid properties (e.g. mass density, viscosity, and thermal conductivity) hence the wall heat flux and wall shear stress.
Thus, without the proper understanding, interpretation of e.g. rheology measurements may fail to give a correct assessment of the fluid properties, even for the simplest multicomponent mixtures.
This paper discusses, with basis in Maxwell-Stefan diffusion theory, under which conditions non-constant composition profiles in the boundary layer must be accounted for.
It is shown, by computational example, how this can impact the estimated wall heat flux and wall shear stress.

Using the ideal ternary gas mixture of \citet{Duncan1962} as an example, mathematical modeling is employed to demonstrate how a compositional gradient must be present to balance the temperature gradient, to achieve zero net mass flux.
Comparing simulations with and without diffusion, the expected consequences of the diffusion-induced, non-constant composition profiles on wall heat flux and wall shear stress is estimated.
 
\section{Mathematical models}
We are considering a single-phase fluid mixture consisting of a set of $N$ unique, distinguishable, inert species. 
It is assumed that each species field, hence the fluid itself, can be modeled as a continuum. 
This implies that molecular scale effects are neglected, and  species and fluid properties are well defined, continuously varying physical fields throughout the fluid domain.
Furthermore, it is assumed homogeneous mixing in the sense that local species properties are taken as volume averages over infinitesimal volumes. 
These assumptions allow the utilization of differential calculus in deriving local governing equations for the species transport.
For a discussion of fluid dynamics at length scales below the thermodynamic limit, see e.g. \citep{Dyson05,Dyson08}.

\subsection{Governing Equations}\label{sec:gov_eq}
The set of steady-state governing equations consists of the Advection-Diffusion equation for each species, the fluid mixture momentum and energy equations, and the restriction that the mass- and mole-fractions must sum to unity:
\begin{gather}
	\bnabla\bcdot\left(\rho_fx_i\bVec{u}_f\right) + \bnabla\bcdot\bVec{j}_{d,i} = 0~;\\
	\bnabla\bcdot\left(\rho_f\bVec{u}_f\bVec{u}_f\right) = -\bnabla P + \bnabla\bcdot\bTen{\tau} + \rho_f\bVec{f}~;\label{eq:momentumeq}\\
	\bnabla\bcdot\left(\rho_fh_{sens,f}\bVec{u}_f\right) =
	 \bnabla\bcdot\left(k_f\bnabla T\right) - \bnabla\bcdot\left(\sum_{i=1}^N{\bVec{j}_{d,i}h_{sens,i}}\right)~;\\
	 \sum_{i=1}^N{x_i}=\sum_{i=1}^N{z_i}=1~.\label{eq:sumtounity}
\end{gather}
$\rho_f$ and $k_f$ are the mass density and thermal conductivity of the fluid, respectively, $x_i$ and $z_i$ are the mass and mole-fractions of species $i$, respectively, and $\bVec{u}_f$ is the mass-averaged advective fluid velocity vector.
The diffusive mass flux is defined as the difference between the convective and advective mass fluxes,
\begin{equation}\label{eq:massflux}
	\bVec{j}_{d,i}\equiv\rho_i\left(\bVec{u}_i-\bVec{u}_f\right)~,
\end{equation}
where $\rho_i$ and $\bVec{u}_i$ are the mass concentration and convective velocity vector of species $i$, respectively.
$P$ and $T$ are pressure and temperature, respectively, $\bTen{\tau}$ is the shear stress tensor, $\bVec{f}$ is the net specific body force vector, and $h_{sens,i}$ is the specific sensible enthalpy of species $i$.

Introducing turbulence, dimensionless variables marked with superscript $+$ (see Appendix), and appropriate simplifications in the near-wall region, the simplified governing equations are obtained:
\begin{gather}
	\left(\nicefrac{\nu _t^+\rho_f^+}{Sc_t}\right)\partial_\bot x_i - j_{d,i,\bot}^+ = 0~;\label{eq:SimpADE}\\
	\rho_f^+\partial_\bot u_{f,\|}^+ = \nicefrac{1}{\left(\nu_f^+ + \nu _t^+\right)}~;\\
	\partial_\bot \left[k_t^+ \left(\partial_\bot \ln c_{P,f}^+\right) T^+ + \left(k_f^++k_t^+\right)\partial_\bot T^+ \right] = 0~.\label{eq:SimpEnEq}
\end{gather}
The $\bot$ and $\|$ indicate the directions normal to the wall and parallel with the wall and bulk flow direction, respectively, and $\partial_\bot$ denotes the dimensionless gradient component in the direction perpendicular to the wall.
$Sc_t$ denotes the turbulent Schmidt number, $\nu_f$ and $\nu_t$ are the fluid and turbulent kinematic viscosities, respectively, $k_t$ is the turbulent thermal conductivity, and $c_{P,f}$ is the fluid specific heat capacity.
It is noted that equations \ref{eq:SimpADE}-\ref{eq:SimpEnEq} are valid in the laminar limit where $\nicefrac{\nu _t^+}{Sc_t}\to0$, $\nu _t^+\to0$, and $k _t^+\to0$.
For more details, refer to \citet{Johnsen15}.

\subsection{Diffusion flux}\label{sec:diff_flux}
Assuming local equilibrium \citep[see][Ch. 3.5]{Kjelstrup08}, local entropy production due to fluid species migration must be positive.
Hence, diffusion acts to minimize the physico-chemical potential of the fluid.
Following \citet{deGroot2011}, the gradient of the dimensionless physico-chemical potential of species $j$ is expressed as $\bnabla\psi_j^{+}=\bnabla\mu_j^{+} - \left(\nicefrac{M_{m,j}}{\UGC T}\right)\bVec{f}_j$, 
where $\mu_j^+=\mu_j^{0+}\left(T,P\right)+\ln{\left(\gamma_jz_j\right)}$, $\gamma_j$, and $M_{m,j}$  are the dimensionless chemical potential, activity coefficient, and molar mass of species $j$, respectively, $\UGC$ is the universal gas constant, and $\bVec{f}_j$ is the net conservative specific external body force acting on species $j$ (e.g. gravitational, centrifugal, and electromagnetic forces per unit mass).
The diffusion mass flux is expressed as a resultant of the scaled, negative physico-chemical potential gradients of all the species in the fluid mixture;
\begin{equation}\label{eq:diffmassflux1}
	j_{d,i,\bot }^+=-\rho_f^+\Ten D_{ij}^+\partial_\bot\psi_j^{+}~,
\end{equation}
where Einstein summation is employed and $D_{ij}$ is the multicomponent diffusion coefficient of species $i$ in species $j$.
It follows from \Eq{massflux} that the diffusive mass fluxes sum to zero, and by combining Eqs. (\ref{eq:SimpADE}) and (\ref{eq:diffmassflux1}), it is seen that zero net mass transport can only be ensured by requiring that the diffusion flux is balanced by turbulent diffusion;
\begin{equation}
	D_{ij}^+\partial_\bot\psi_j^+ = -\left(\nicefrac{\nu_t^+}{Sc_t}\right)\partial_\bot x_i~.
\end{equation}

Mechanical equilibrium is assumed throughout the paper.
\Eq{momentumeq} then reduces to $\bnabla P=-\rho_f\sum_{i=0}^Nx_i\bVec{f}_i$, where it has been assumed that the total, specific, external body force acting on the fluid can be decomposed into species-specific components, $\bVec{f}_i$.
Expanding the chemical potential gradient in its partial derivatives, \Eq{diffmassflux1} may now be written as \citep{Taylor93,Kocherginsky2016}
\begin{multline}\label{eq:diffmassflux2}
	j_{d,i,\bot }^+=-\rho_f^+ \Ten D_{ij}^+\Bigg[\Ten\Gamma_{jk}\Ten\Lambda_{kl}\partial_\bot x_l + \Ten d_{T,j}\partial_\bot\ln{\left(T^++T_{wall}^{0+}\right)} \\
	+	\frac{1}{\UGC T}\left(V_{m,j}-\frac{M_{m,j}}{\rho_f}\right)\partial_\bot P - \frac{1}{\UGC T}\frac{\nu_{f,wall}}{u_\tau}\left(F_{j,\bot} - M_{m,j}\sum_{k=1}^Nx_kf_{k,\bot}\right)\Bigg]~,
\end{multline}
where the diffusiophoretic and thermophoretic driving force coefficients are defined as ${\Gamma }_{jk} = \nicefrac{\partial\mu_j^+}{\partial z_k}$ and $d_{T,j} = \nicefrac{\partial\mu_j^+}{\partial\ln{\left(T^++T_{wall}^{0+}\right)}}$, respectively, $\Lambda_{kl}\partial_\bot X_l=\partial_\bot z_k$, the subscript $wall$ indicates property values at the wall, $V_m$ is the molar volume, $u_\tau$ is the shear velocity, $F_{j,\bot}=M_{m,j}f_{j,\bot}$ (summation not intended) is the molar body force acting on species $j$ in the wall-normal direction, and it is implicitly understood that partial derivatives are taken with respect to one variable while keeping all other variables constant.
Hence, the diffusion flux can be decomposed into a diffusiophoretic term due to composition gradients, a thermophoretic term due to temperature gradients, a barophoretic term due to pressure gradients, and a forced diffusion term due to external forces.
It is noted that $\partial_\bot P=0$ if $\sum_{k=1}^Nx_kf_{k,\bot}=0$.

\section{How can Non-Constant Composition Profiles Occur in the Boundary Layer?}
Based on the governing equations and relations described in the previous section, this section discusses the mathematical requirements for obtaining non-constant composition profiles in the boundary-layer.
Three main regimes are discussed:
\begin{enumerate}
	\item Isothermal conditions in the absence of body forces.
	\item Non-isothermal conditions in the absence of body forces.
	\item Isothermal conditions under influence of body forces (gravitational/centrifugal).
\end{enumerate}
The discussion is of a general nature, and it is shown how all three scenarios allow (at least mathematically), or even demand the generation of non-constant composition profiles.
Due to the additive nature of the diffusive driving forces, the scenarios discussed cover most situations that can occur.
Body forces due to electromagnetic forces on charged or polarized species has not been discussed due to the difficulties associated with drawing general conclusions regarding the resulting composition profiles.
This difficulty is due to the multitude of different situations that can arise depending on the electric charge of the species.
For instance, in a given circumstance, the mass averaged electrostatic body force can be zero, causing the pressure gradient to vanish.
Replacing one of the species with another species of different electric charge, will change this, producing a non-zero pressure gradient.
Moreover, separation of electrically charged species will cause deviation from quasi neutrality of the fluid.
Hence, the Poisson equation, for the electrostatic field, must be added to the set of governing equations to calculate local electrostatic field strengths.
Furthermore, electrically charged species tend to be reactive, not inert.
This paper discusses inert fluids, only.
That being said, there is no doubt, however, that electromagnetic forces, similar to gravitational/centrifugal forces, can be the source of species separation.

\subsection{Isothermal conditions in the absence of body forces}\label{sec:isothermal}
\Eq{SimpADE} can be written as a homogeneous system of equations:
\begin{equation}\label{eq:isothermalADE}
	D_{X,il}^+\partial_\bot x_l=0~,
\end{equation}
where $D_{X,il}^+ = \left[\left(\nicefrac{\nu _t^+}{Sc_t}\right)\delta_{il} + D_{ij}^+\Gamma_{jk}\Lambda_{kl}\right]$ and $\delta_{il}$ is the Kronecker delta.
\Eq{isothermalADE} has a non-trivial solution ($\partial_\bot x_l\neq0$) if and only if $-\nicefrac{\nu_t^+}{Sc_t}$ is a non-zero eigenvalue of the matrix product $\bTen{D}^+\bTen{\Gamma}\bTen{\Lambda}$.
Mathematically this is permitted, but it would be an absolutely remarkable coincidence if this was to be true for all wall distances in a physical boundary layer.
Hence, it can be deduced that it is very unlikely that a non-constant composition profile can develop in the absence of temperature gradients and/or body forces.
It is noted that at the wall, where $\nicefrac{\nu_t^+}{Sc_t}\to0$, the required condition for a non-trivial solution reduces to $\det{\left(\bTen\Gamma\right)}=0$, since both $\bTen D^+$ and $\bTen\Lambda$ are invertible.
In particular, this is impossible for ideal mixtures ($\gamma_j=1~\forall~j$), since $\bTen\Gamma$ becomes diagonal with non-zero elements, only.

\subsection{Non-isothermal conditions in the absence of body forces}\label{sec:nonisothermal}
In the presence of non-zero temperature derivatives in the wall-normal direction, there must be a balance between the turbulent-diffusiophoretic diffusion on one side and thermophoretic diffusion on the other.
This balance can be formulated as the non-homogeneous system of equations
\begin{equation}\label{eq:nonisothermalADE}
	D_{X,il}^+\partial_\bot x_l=-D_{ij}^+ d_{T,j}\partial_\bot \ln \left(T^+ + T_{wall}^{0+}\right)~.
\end{equation}
It is evident that non-zero mass-fraction gradients are required to counter the thermophoresis, in general.
At the wall, \Eq{nonisothermalADE} reduces to
\begin{equation}\label{eq:idealwallcond}
	\left.\Gamma_{jk}\partial_\bot z_k\right|_{wall} = -d_{T,j,wall}\nicefrac{\Pran_{wall}}{T_{wall}^{0+}} ~,
\end{equation}
which in combination with \Eq{sumtounity} produces the requirement that $\sum_{j=1}^N{z_jd_{T,j}}=0$ holds at the wall, for ideal mixtures.
This implies that $\Ten d_{T}$ components of both positive and negative values must exist at the wall, for ideal mixtures.

\subsection{Isothermal conditions under influence of body forces}
In the presence of external body forces in the wall-normal direction, there must be a balance between the turbulent-diffusiophoretic, barophoretic, and forced diffusion.
From Eqs. \ref{eq:SimpADE} and \ref{eq:diffmassflux2}, we have
\begin{equation}\label{eq:forceddiff}
	\Ten D_{X,il}^+\partial_\bot x_l = -\frac{1}{\UGC T}\Ten D_{ij}^+\left[\left(V_{m,j}-\frac{M_{m,j}}{\rho_f}\right)\partial_\bot P - \frac{\nu_{f,wall}}{u_\tau}\left(F_{j,\bot} - M_{m,j}\sum_{k=1}^Nx_kf_{k,\bot}\right)\right]~.
\end{equation}
If the body forces are due to gravity or centrifugal forces, the forced diffusion term cancels out due to the non-discriminating effect on the various species, $f_{j,\bot} \equiv f_\bot$, where
\begin{equation}
	f_\bot = \begin{cases}
		g_\bot &\text{for gravitational body force},\\
		\omega^2 r_\bot &\text{for centrifugal body force},
	\end{cases}
\end{equation}
where $r_\bot$ is the wall-normal distance from the center of rotation.
\Eq{forceddiff} thus reduces to the non-homogeneous system of equations
\begin{equation}\label{eq:forceddiff2}
	\Ten D_{X,il}^+\partial_\bot x_l = \Ten D_{ij}^+\left[\frac{\nu_{f,wall}}{u_\tau}\frac{f_\bot M_{m,j}}{\UGC T}\left(1-\frac{\rho_f}{\rho_j^0}\right)\right]~,
\end{equation}
where $\rho_j^0=\nicefrac{M_{m,j}}{V_{m,j}}$  denotes the mass density of the pure species, $j$, and summation over the repeated $j$ index inside the square brackets is not intended.
If the mass density of the pure species differs from that of the fluid mixture, it is evident that non-constant composition profiles are required to counter the effect of the forced diffusion.

For an ideal mixture of perfect gases in the absence of turbulence ($\nicefrac{\nu_t^+}{Sc_t}=0$ everywhere), two special cases are emphasized, namely 1) dilute mixture, and 2) binary mixture.
For a mixture dilute in species $i$, $1-z_i\approx1$, and the solution to \Eq{forceddiff2} becomes
\begin{equation}
	z_i = z_{i,bulk}\exp\left[-\frac{\nu_{f,wall}}{u_\tau}\frac{f_\bot\left(M_{m,i}-\bar M_{m,i}^*\right)}{\UGC T}\left(y_{bulk}^+-y^+\right)\right]~,
\end{equation}
where subscript $bulk$ indicates property values at a reference point at the outskirt of the boundary layer, $\bar M_{m,i}^*\equiv\tfrac{1}{\left(1-z_i\right)}\sum_{\substack{j=1,\\ j\neq i}}^N M_{m,j}z_j$, it was assumed that $f_\bot$ is approximately constant throughout the boundary layer, and $y^+$ is the dimensionless distance to the wall.
Similarly, for a binary mixture, it can be shown that the ratio between the mole-fractions of the two species is given by
\begin{equation}
	\left(\frac{z_1}{z_2}\right) = \left(\frac{z_{1,bulk}}{z_{2,bulk}}\right)\exp\left[-\frac{\nu_{f,wall}}{u_\tau}\frac{f_\bot\left(M_{m,1}-M_{m,2}\right)}{\UGC T}\left(y_{bulk}^+-y^+\right)\right]~.
\end{equation}
In both cases, segregation is governed by the dimensionless group $\frac{\nu_{f,wall}}{u_\tau}\frac{f_\bot\Delta M_m}{\UGC T}$, where $\Delta M_m$ denotes the difference between the species' molar mass and the representative molar mass of the other species.
In particular, $\left|\frac{\UGC T}{f_\bot\Delta M_m}\right|$ is the segregation length.
For the segregation length to be of the same order of magnitude as the boundary layer thickness, a large $f_\bot$ is generally required, many orders of magnitude larger than the gravitational force.
The effect of turbulence will be to reduce the magnitude of the composition  derivatives, since it contributes with a positive addition in $\Ten D_{X,il}^+$ on the left hand side of \Eq{forceddiff2}.
Hence, the effect of turbulence will be reduced segregation efficiency.

\section{Ternary Ideal Mixture - Computed example}
\subsection{Model fluid}
The model fluid is a ternary, calorically perfect mixture of perfect gasses consisting of $50$, $25$, and $25~\text{mass-\%}$ of $H_2$, $N_2$, and $CO_2$, respectively.
Species specific heat capacities were extracted from \citep{NISTChem} while species specific viscosities and thermal conductivities were calculated based on Lennard-Jones parameters found in \citep{Anderson06}.
Details regarding the modeling of species and mixture material properties (mass density, viscosity, etc.) can be found in \citep{Johnsen15}.
Species specific input data are summarized in \tab{MatProp}.

	\begin{table}
  		\begin{center}
			\def~{\hphantom{0}}
			\begin{tabular}{lllll}
	&       &       & \multicolumn{2}{c}{Lennard-Jones param.} \\
	& \multicolumn{1}{c}{$M_m$} & \multicolumn{1}{c}{$c_P^1$} & \multicolumn{1}{c}{$d$} & \multicolumn{1}{c}{$\Omega^1$} \\
	& \multicolumn{1}{c}{$[\nicefrac{\mathrm{kg}}{\mathrm{mol}}]$} & \multicolumn{1}{c}{$[\nicefrac{\mathrm{J}}{\mathrm{molK}}]$} & \multicolumn{1}{c}{$[\mathrm{\text{\AA}}]$} & \multicolumn{1}{c}{$[-]$} \\
	\midrule
	$H_2$ & \multicolumn{1}{l}{0.002016} & \multicolumn{1}{c}{28.84} & \multicolumn{1}{c}{2.915} & \multicolumn{1}{c}{0.857} \\
	$N_2$ & \multicolumn{1}{l}{0.02801} & \multicolumn{1}{c}{29.12} & \multicolumn{1}{c}{3.681} & \multicolumn{1}{c}{1.022} \\
	$CO_2$ & \multicolumn{1}{l}{0.04401} & \multicolumn{1}{c}{37.12} & \multicolumn{1}{c}{3.996} & \multicolumn{1}{c}{1.296} \\
	\midrule
	\multicolumn{5}{l}{$^1$ values at $298\mathrm{K}.$} \\
\end{tabular}%
	  		\caption{Species specific properties.}
	  		\label{tab:MatProp}
  		\end{center}
	\end{table}

For a ternary mixture, there are two independent mass-fraction equations in addition to the velocity and temperature equations.
Moreover, there are only two independent diffusive mass-fluxes, and the matrices that take part in \Eq{diffmassflux2} are $2\times2$ matrices.

The elements of the diffusivity matrix, $\bTen{D}$, can be expressed as \citep{Taylor93}
\begin{equation}
\begin{aligned}
	\Ten D_{11}&=\nicefrac{\Dbar_{13}\left[z_1\Dbar_{23} + \left(1 - z_1\right)\Dbar_{12}\right]}{S}~,\\ 
  \Ten D_{12}&=\nicefrac{z_1\Dbar_{23}\left[\Dbar_{13} - \Dbar_{12}\right]}{S}~,\\ 
  \Ten D_{21}&=\nicefrac{z_{2}\Dbar_{13}\left[\Dbar_{23} - \Dbar_{12}\right]}{S}~,\\ 
  \Ten D_{22}&=\nicefrac{\Dbar_{23}\left[z_2\Dbar_{13} + \left(1 - z_2\right)\Dbar_{12}\right]}{S}~,
  \end{aligned}
\end{equation}
where $S={{z}_{1}}{{\Dbar}_{23}}+{{z}_{2}}{{\Dbar}_{13}}+{{z}_{3}}{{\Dbar}_{12}}$, the $\Dbar_{ij}$ are the binary Maxwell-Stefan diffusion coefficients, and the indexes $1,2$ and $3$ relate to $H_2$, $N_2$, and $CO_2$, respectively. 
The binary Maxwell-Stefan diffusivities employed by \citet{Duncan1962} are cited in \tab{Dbarval}.
It is noted that the Onsager reciprocal relation implies that $\Dbar_{ij}=\Dbar_{ji}$ \citep{Hirschfelder64,Muckenfuss73}.

	\begin{table}
  		\begin{center}
			\def~{\hphantom{0}}
			\begin{tabular}{ll}
    \toprule
    $H_2 - N_2$ & $\Dbar_{12}=8.33\cdot10^{-5}\nicefrac{\mathrm{m^2}}{\mathrm{s}}$ \\
    $H_2 - CO_2$ & $\Dbar_{13}=6.80\cdot10^{-5}\nicefrac{\mathrm{m^2}}{\mathrm{s}}$ \\
    $N_2 - CO_2$ & $\Dbar_{23}=1.68\cdot10^{-5}\nicefrac{\mathrm{m^2}}{\mathrm{s}}$ \\
    \bottomrule
 \end{tabular}%
	  		\caption{Binary Maxwell-Stefan diffusion coefficients for the ternary $H_2-N_2-CO_2$ mixture \citep{Duncan1962}.}
	  		\label{tab:Dbarval}
  		\end{center}
	\end{table}

\citet{Bogatyrev2015} reported thermal diffusion factors, $\alpha_{T}$, as functions of composition for each of the mixture species.
The thermal diffusion factors are related to the diffusiophoretic driving force coefficients, $d_{T,j}$ via the thermal diffusion ratio, $\Ten k_{T,k}$, by \citep[see][]{vandervalk63}
\begin{equation}\label{eq:dTj}
	\Ten d_{T,j} = \Ten\Gamma_{jk}\Ten k_{T,k}~,
\end{equation}
where
\begin{equation}
	k_{T,k} = z_k\sum_{\substack{l=1 \\ l\neq k}}^N{z_l\alpha_{T,kl}}~.
\end{equation}
For ideal mixtures, \Eq{dTj} reduces to
\begin{equation}\label{eq:dTjideal}
	d_{T,j} = \sum_{\substack{l=1 \\ l\neq j}}^N{z_l\alpha_{T,jl}}~.
\end{equation}
Using the experimental data points at $z_l\approx0.5$ from \citet{Bogatyrev2015}, the thermal diffusion factors and thermophoretic driving force coefficients given in \tab{thermaldiffratio} were obtained.
For simplicity, constant $d_{T,j}$ were employed in the simulations.

	\begin{table}
  		\begin{center}
			\def~{\hphantom{0}}
			\begin{tabular}{lcccc}
	\boldmath{}\textbf{$j$}\unboldmath{} & \boldmath{}\textbf{$\alpha_{T,j,H_2}$}\unboldmath{} & \boldmath{}\textbf{$\alpha_{T,j,N_2}$}\unboldmath{} & \boldmath{}\textbf{$\alpha_{T,j,CO_2}$}\unboldmath{} & \boldmath{}\textbf{$d_{T,j}$}\unboldmath{} \\
	\midrule
	$H_2$ 	&       & 0.32  & 0.38  & 0.161 \\
	$N_2$ 	& 0.24  &       & 0.06  & 0.073 \\
	\bottomrule
\end{tabular}%
	  		\caption{Thermal diffusion factors, $\Ten\alpha_{T,jl}$, based on data from \citep{Bogatyrev2015} and resulting thermophoretic driving force coefficients, $\Ten d_{T,j}$ (assuming ideal mixture, see \Eq{dTjideal}).}
	  		\label{tab:thermaldiffratio}
  		\end{center}
	\end{table}

\subsection{Simulation setup}
The equations were solved in a numerical modeling framework described by \citet{Johnsen15}.
The simulations assume fully developed turbulent flow parallel to the wall.
Moreover, it is assumed that gradients in the main flow direction are negligible and that gradients perpendicular to the wall vanish in the bulk.
Additional details can be found in \citep{Johnsen15}.

The wall and bulk temperatures were set equal to the \cite{Bogatyrev2015} temperatures of $280\mathrm{K}$ and $800\mathrm{K}$, respectively, and a range of bulk flow velocities were employed.
The boundary conditions employed in the simulations are summarized in \tab{BC}.

The simulations were conducted on a 1-dimensional computational mesh consisting of $30$ grid points logarithmically distributed between the wall and the bulk.
A grid sensitivity study showed that the wall heat flux and wall shear stress varied with less than 1\% between a grid with 30 grid points and one with 100 grid points.
The first grid point was located $10^{-7}\mathrm{m}$ away from the wall, and the bulk node was located $10^{-3}\mathrm{m}$ away from the wall.
The results were insensitive to decreasing the first node distance to the wall.

To isolate the effect of non-zero composition gradients, simulations with and without diffusion were conducted.
In the simulations without diffusion, the multicomponent diffusion coefficients were set to zero, $D_{ij}=0~\forall~i,j$.

	\begin{table}
  		\begin{center}
			\def~{\hphantom{0}}
			\begin{tabular}{llll}
    \textbf{Boundary Condition} & \textbf{Variable} & \textbf{Value} & \textbf{Unit} \\
    \midrule
    bulk mass-fractions & $X_{H_2,bulk}$ & 0.5   & $\nicefrac{\mathrm{kg}}{\mathrm{kg}}$ \\
          & $X_{N_2,bulk}$ & 0.25  & $\nicefrac{\mathrm{kg}}{\mathrm{kg}}$ \\
    wall diffusion mass flux & $j_{d,H_2,\bot,wall}$ & 0     & $\nicefrac{\mathrm{kg}}{\mathrm{m^2s}}$ \\
          & $j_{d,N_2,\bot,wall}$ & 0     & $\nicefrac{\mathrm{kg}}{\mathrm{m^2s}}$ \\
    bulk temperature & $T_{bulk}$ & 800   & $\mathrm{K}$ \\
    wall temperature & $T_{wall}$ & 280   & $\mathrm{K}$ \\
    bulk flow velocity & $u_{f,\|,bulk}$ & 1,2,5,10 & $\nicefrac{\mathrm{m}}{\mathrm{s}}$ \\
    \bottomrule
\end{tabular}%
	  		\caption{Boundary conditions employed in simulations.}
	  		\label{tab:BC}
  		\end{center}
	\end{table}

\subsection{Simulation results}
In the simulations without diffusion, the mass-fraction profiles were constant throughout the boundary layer, and fluid properties varied only due to the varying temperature.
In simulations including diffusion, however, non-constant mass-fraction profiles resulted to balance the thermophoretic diffusion by turbulent-diffusiophoretic diffusion, to maintain zero net diffusive transport.
The resulting mass-fraction profiles are shown in \fig{MassFrac}, for the various bulk flow velocities (darker curve corresponds to higher velocity).
Generally, the mass-fraction of $CO_2$ increased towards the wall while $H_2$ and $N_2$ mass-fractions decreased.
Due to the fluid properties- (e.g. mass density and viscosity) composition dependency, the simulations predict a bulk flow velocity dependency in these.

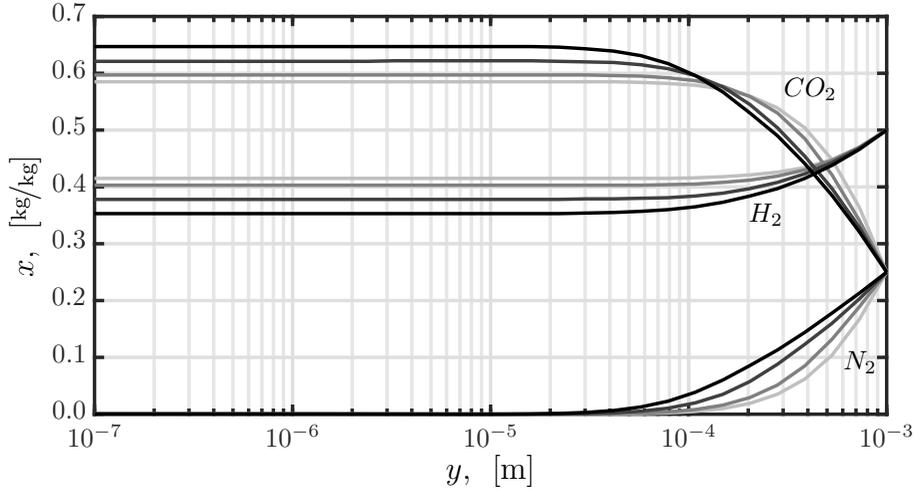
\InsPSfragFig{Figure1}{Calculated mass-fractions, $x$, plotted against wall distance, $y$, for the three species $H_2$, $N_2$, and $CO_2$, for the bulk flow velocities $1$ (light grey), $2$, $5$, and $10\nicefrac{\mathrm{m}}{\mathrm{s}}$ (black).}{MassFrac}

In figures (\ref{fig:DenWall}) and (\ref{fig:ViscWall}), respectively, the wall mass density and dynamic viscosity are shown as functions of the bulk flow velocity.
It is seen that the failure to consider diffusion, resulted in underprediction of the fluid mass density and viscosity at the wall.
Moreover, in the absence of diffusion, the mass density and viscosity are insensitive to the flow velocity since the wall temperature was fixed.
However, due to their dependence on composition, they become flow velocity dependent when diffusion is considered.
The near-wall enrichment of $CO_2$, which is the densest fluid component, causes the fluid density to increase towards the wall.
Since the enrichment increases with the flow velocity (see \fig{MassFrac}), the mass density at the wall also increases with flow velocity.

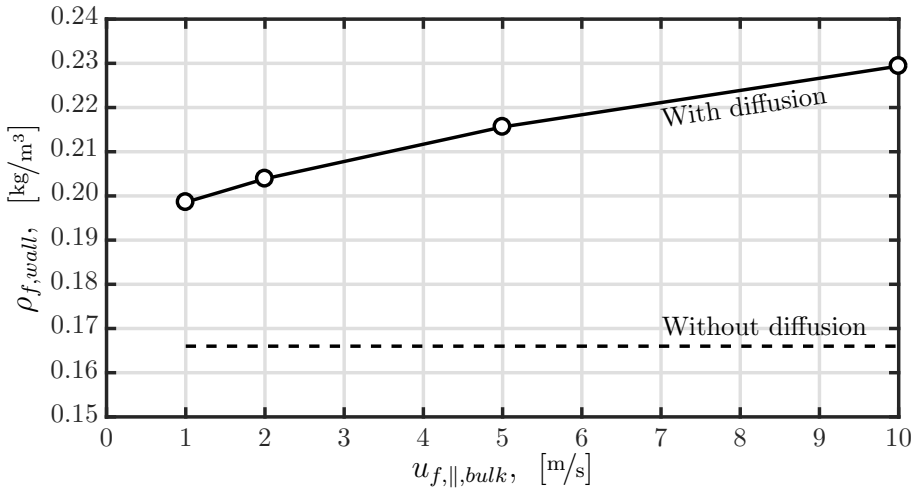
\InsPSfragFig{Figure2}{Calculated fluid mass density at the wall, $\rho_{f,wall}$, plotted against bulk flow velocity, $u_{\|,bulk}$, with and without diffusion included.}{DenWall}

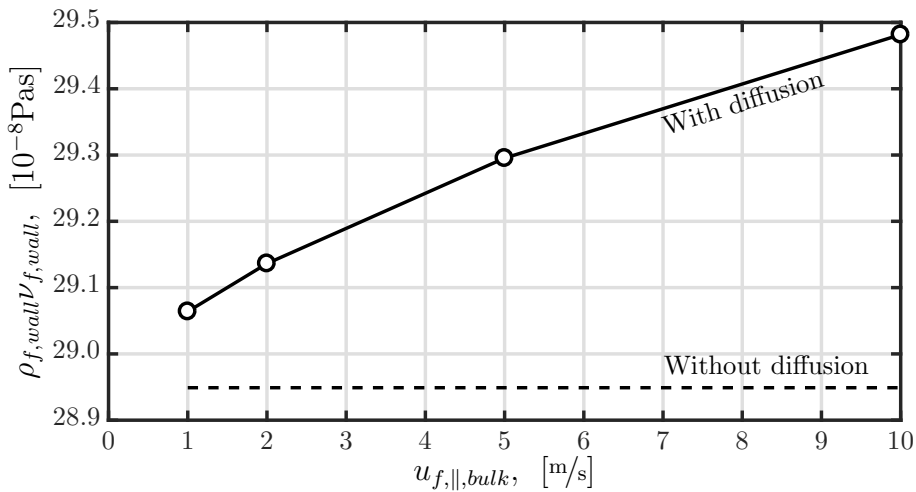
\InsPSfragFig{Figure3}{Calculated dynamic fluid viscosity at the wall, $\rho_{f,wall}\nu_{f,wall}$, plotted against bulk flow velocity, $u_{\|,bulk}$, with and without diffusion included.}{ViscWall}

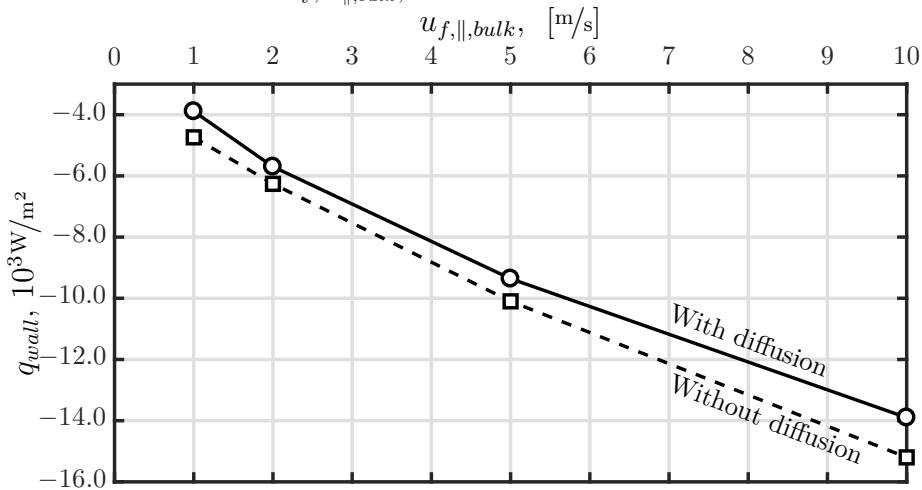
\InsPSfragFig{Figure4}{Calculated wall heat flux, $q_{wall}$, plotted against bulk flow velocity, $u_{\|,bulk}$, with and without diffusion included (negative heat flux indicates that heat is flowing from the fluid into the wall).}{QWall}

In \fig{QWall}, the wall heat flux is shown as a function of the bulk flow velocity, for simulations with and without diffusion.
Negative heat flux indicates that the heat flows from the fluid into the wall, and the magnitude of the wall heat flux generally increases with the bulk flow velocity, as expected, due to the increased efficiency of turbulent heat transport.
The calculated wall heat flux was $18$ and $8.6\%$ lower with diffusion than without, in the $u_{f,\|,bulk}=1$ and $10\nicefrac{\mathrm m}{\mathrm s}$ simulations, respectively.
This was mainly due to the predicted diffusion-induced reduction of the fluid's thermal conductivity at the wall.

In \fig{TauWall}, the wall shear stress is shown as a function of the bulk flow velocity, for simulations with and without diffusion.
The wall shear stress increases with increasing flow velocity, as expected, and the simulations predict that diffusion will increase the growth rate.
The calculated wall shear stress was $4.7$ and $12\%$ higher with diffusion than without, in the $u_{f,\|,bulk}=1$ and $10\nicefrac{\mathrm m}{\mathrm s}$ simulations, respectively.
This was mainly due to the diffusion-induced increase in viscosity.

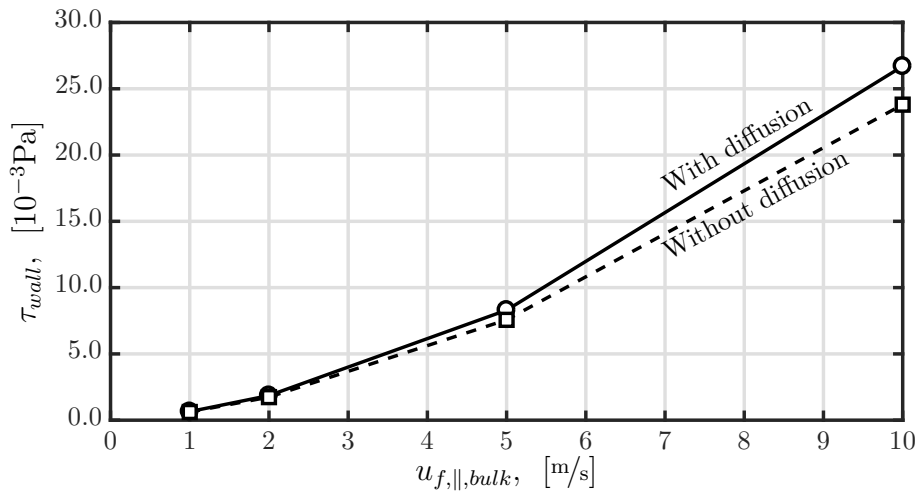
\InsPSfragFig{Figure5}{Calculated wall shear stress, $\tau_{wall}$, plotted against bulk flow velocity, $u_{\|,bulk}$, with and without diffusion included.}{TauWall}

\section{Conclusion}
Employing mathematical modeling, it has been shown that the combined turbulent, diffusiophoretic and thermophoretic diffusion can have a considerable effect on composition profiles in the boundary-layer for inert, multicomponent fluids.
This is of importance for e.g. the interpretation of measurements relying on wall heat flux and/or wall shear stress data. 
Fluid properties derived from such measurements (e.g. viscosity or thermal conductivity) typically depend on the fluid composition, which may differ significantly between the bulk and the wall.
Hence, failing to take diffusion into account may result in interpreted fluid properties that represent the bulk fluid inaccurately.
Moreover, the effect of non-constant composition profiles may impact the design and mathematical modeling of flow systems sensitive to drag and/or heat exchange.

Mathematical proof was given to support the following statements for inert mixtures:
\begin{itemize}
	\item Under isothermal conditions, in the absence of body forces: 
	\begin{itemize}
		\item Non-constant composition profiles requires that $-\nicefrac{\nu_t^+}{Sc_t}$ is an eigenvalue of the matrix product $\bTen{D}^+\bTen{\Gamma}\bTen{\Lambda}$.
		\item Non-zero compositional gradients at the wall requires that $\det{\left(\bTen{\Gamma}\right)}=0$.
		\item Ideal mixtures are not permitted to have non-zero compositional gradients at the wall.
	\end{itemize}
	\item Under non-isothermal conditions, in the absence of body forces:
	\begin{itemize}
		\item Non-constant composition profiles are required to counter thermophoresis.
		\item For ideal mixtures, the thermophoretic driving force coefficients must obey $\sum_{j=1}^N{z_jd_{T,j}}=0$.
	\end{itemize}
	\item Under isothermal conditions influenced by gravitational/centrifugal body forces:
	\begin{itemize}
		\item Non-constant composition profiles are required to balance barophoresis if the pure state mass density of one or more of the species differ from the fluid mass density.
		\item Under laminar conditions, the segregation length is given by $\left|\nicefrac{\left(\UGC T\right)}{\left(f_\bot\Delta M_m\right)}\right|$ in diluted an binary mixtures.
		\item Turbulence will reduce the segregation efficiency, hence increase the segregation length.
	\end{itemize}
\end{itemize}

\vspace{0.5cm}
Gratitude goes to all the colleagues at the research group of flow technology at dept. Process technology, SINTEF Industry, in Trondheim, Norway.
Without the vibrant research environment and fruitful discussions, this paper would not be.
In particular, Roar Meland and John C. Morud contributed with to-the-point comments and discussions and with reviewing the paper.
The study was financed by SINTEF Industry and NTNU.

\appendix
\section{Dimensionless Variables}
The model equations presented in this paper are presented in dimensionless form.
Dimensionless variables are denoted by superscript $+$. 
When making the conservation equations dimensionless, typical wall unit scaling is employed.
The subscript $wall$ indicate property values at the wall.
Selected scaled variables are given below. 

The shear velocity is defined as
\begin{equation}
	u_\tau=\sqrt{\nicefrac{\tau_w}{\rho_{f,wall}}}~,
\end{equation}
the dimensionless wall distance is defined as
\begin{equation}
	y^+=\nicefrac{yu_\tau}{\nu_{f,wall}}~,
\end{equation}
where $y$ is the normal distance to the wall.
$\nu_{f,wall}=\nicefrac{\mu_{f,wall}}{\rho_{f,wall}}$ is the kinematic viscosity at the wall, the dimensionless fluid velocity is defined as
\begin{equation}
	u_f^+=\nicefrac{u_f}{u_\tau}~,
\end{equation}
and the dimensionless mass flux is given by
\begin{equation}
	j^+=\nicefrac{j}{\rho_{f,wall}u_\tau}~.
\end{equation}
Fluid properties are typically converted to wall units by scaling with the value at the wall, e.g.: 
\begin{align}
	\text{mass density:\quad\quad}&\rho_f^+=\nicefrac{\rho_f}{\rho_{f,wall}}~;\\
	\text{kinematic viscosity:\quad\quad}&\nu_f^+=\nicefrac{\nu_f}{\nu_{f,wall}}~;\\
	\text{thermal conductivity:\quad\quad}&k_f^+=\nicefrac{k_f}{k_{f,wall}}~.
\end{align}
The dimensionless, turbulent thermal conductivity is defined as
\begin{equation}
	k_t^+=\nu_t^+\rho_f^+c_{P,f}^+\left(\nicefrac{\Pran_{wall}}{\Pran_t}\right)~,
\end{equation}
and the dimensionless turbulent kinematic viscosity is modeled as \citep{Johansen91}
\begin{equation}
\nu _{t,f}^{+}=\begin{cases}
	\left(\nicefrac{y^+}{11.15}\right)^3&\text{for $y^+<3.0$},\\
	\left(\nicefrac{y^+}{11.4}\right)^2 - 0.049774&\text{for $3.0\le y^+\le 52.108$},\\
	0.4y^+&\text{for $52.108<y^+$}.
\end{cases}
\end{equation}
Diffusivities are scaled by the fluid kinematic viscosity at the wall, e.g.
\begin{equation}
	D_{ij}^+ = \nicefrac{D_{ij}}{\nu_{f,wall}}~.
\end{equation}
The dimensionless temperature is given by
\begin{equation}
	T^+=\nicefrac{Tu_\tau\rho_{f,wall}c_{P,f,wall}}{q_w} - T_{wall}^{0+}~,
\end{equation}
where
\begin{equation}
	T_{wall}^{0+} = \nicefrac{T_{wall}u_\tau\rho_{f,wall}c_{P,wall}}{q_{wall}}~,
\end{equation}
and $q_{wall}=-\left. k_f\partial_\bot T\right|_{wall}$ is the wall heat flux.
The Prandtl number is given by
\begin{equation}
	\Pran=\nicefrac{c_{P,f}\rho_f\nu_f}{k_f}~.
\end{equation}
Constant turbulent Prandtl and Schmidt numbers of $\Pran_t=0.85$ and $\Sc_t=0.7$, respectively, were employed in simulations.

\newpage
\bibliographystyle{jfm}
\bibliography{References}

\end{document}